\documentclass[12pt]{cernart}
\usepackage{times}
\usepackage{graphicx}
\usepackage{cite}
\begin{document}
\begin{titlepage}


\vspace{1cm}
\title{\Large $\Delta G$ from Compass}

\begin{center}
Krzysztof Kurek\footnote{address: Andrzej Soltan Institute for
Nuclear Studies,
 Hoza 69,00-681 Warsaw, Poland, kurek@fuw.edu.pl,
Work partially supported by SPUB
621/E-78/SPB/CERN/P-03/DWM576/2002-2006.}\\
 \vspace{1cm}
 On behalf of COMPASS Collaboration\\
\vspace{1cm}
 \it {Talk given on XIV International Workshop \\
 on Deep Inelastic Scattering, 20-24 April,2006, Tsukuba, Japan}
 \end{center}

\begin{abstract}
Measurements of the gluon polarization $\frac{\Delta G}{G}$ via
the open charm channel and based on the helicity asymmetry of
large transverse-momentum hadrons in the final state are
presented. The data have been collected in the years 2002-2004 by
the COMPASS experiment at CERN using a 160 GeV/c polarized muon
beam scattered off a polarized $^6$LiD target. The new result for
$\frac{\Delta G}{G}$ from the charm channel is $-0.57 \pm 0.41
(stat.)$ at $x_G \simeq 0.15$ and scale $\mu^2 \simeq 13 $
(GeV/c)$^2$.\\ The gluon polarization from high-$p_T$ hadron pairs
is $\frac{\Delta G}{G} = 0.016\pm 0.058 (stat.)\pm 0.055 (syst.)$
at $x_G \simeq 0.085^{+0.07}_{-0.035}$ ($Q^2 < 1$ (GeV/c)$^2$ and
$\mu^2 \simeq 3$ (GeV/c)$^2$)
\end{abstract}

\end{titlepage}


\section{Introduction}
The EMC spin asymmetry measurement \cite{emc} and the naive
interpretation of the results following of Ellis-Jaffe sum rule
\cite{ej} have introduced the so-called "spin crisis": quarks
carry very small fraction of the nucleon's helicity. The next
experiments at CERN, DESY and SLAC  confirmed that quarks are only
responsible for roughly $1/3$ of nucleon's helicity. The quark
helicity distributions $\Delta q_{i}(x,Q^2)$ are related to
vector-axial quark current which is not conserved due to the
Adler-Bell-Jackiw anomaly. This fact allows to explain the spin
crisis by changing the interpretation of the measurement: instead
of quark spin contents $\Delta \Sigma = \int_0^1
\sum_{i=1}^{n_f}q_i(x,Q^2) dx $ the combination $\Delta \Sigma -
\frac{3 \alpha _s}{2 \pi } \Delta G$ is measured, where $\Delta G$
is a gluon polarization inside the nucleon.  The spin crisis and
the violation of the Ellis-Jaffe sum rule can be then avoided if
$\Delta G$ is large enough. To complete the picture,  beside the
quark's helicity $\Delta \Sigma$, and the gluon polarization
$\Delta G$ also an orbital angular momentum of quarks and gluons
can build the nucleon spin structure. This interpretation was a
"driving force" in preparation a series of new polarized DIS type
experiments related to direct measurement of $\Delta G$:  HERMES
in DESY, SMC and COMPASS at CERN, STAR and PHOENIX at RHIC.\\ In
this paper I will present new results for a direct measurement of
gluon polarization $\frac{\Delta G}{G}$ obtained by COMPASS
collaboration after analyzing the data sets collected in years
2002-2004. The experiment is using a 160 GeV/c polarized muon beam
from SPS at CERN scattered off polarized $^6$LiD target (for more
details see F.Kunne~\cite{fab}). In the LO QCD approximation the
only subprocess which probes gluons inside nucleon is Photon-Gluon
Fusion (PGF).
 There are two ways allowing direct access to gluon
polarization via the PGF subprocess available in the COMPASS
experiment: the open charm channel where the events with
reconstructed $D^0$ mesons are used and the production of two
hadrons with relatively high-$p_T$ in the final state. The open
charm channel guarantees no physical background because the PGF
subprocess is the only possible mechanism for charm quarks pair
production in LO QCD approximation (NLO corrections, the so-called
"intrinsic" charm mechanism as well as resolved photon
contribution are neglected in this analysis). Therefore the
estimation of the gluon polarization in this case is much less
Monte-Carlo (MC) dependent than in the two high-$p_T$ hadrons
method, where the complicated background requires very good MC
description of the data. On the other hand the statistical
precision in high-$p_T$ hadrons method is much higher than in the
open charm channel.

\section{$\frac{\Delta G}{G}$ from open charm channel}

\begin{figure}[t]
\begin{center}
\includegraphics[width=0.9\hsize]{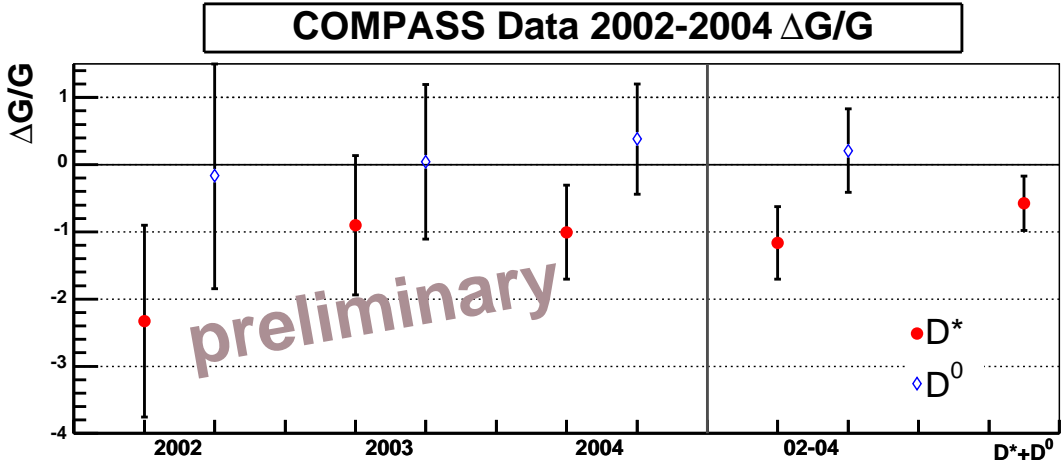}
\caption{The value of$\frac{\Delta G}{G}$ obtained from measured
helicity asymmetry for charm events.}
\end{center}
\label{fig:charm}
\end{figure}

For events with charm quarks production a helicity asymmetry has
been measured. Charm quarks were tagged by measuring  $D$ mesons
in one of the two channels: the $D^0$ meson decaying into the
"golden" channel i.e. to a kaon and a pion, and $D^*$ decaying
into soft pion and a $D^0$ with subsequent decay (so-called
$D^*$-tagged events). For particle identification the RICH
detector was used. The gluon polarization $\frac{\Delta G}{G}$ is
related to the measured helicity asymmetry as follows:
\begin{equation}
A_{LL}= \frac{S}{S+B} a_{LL}\frac{\Delta G}{G}, \label{eq:charm}
\end{equation}
where $S$ and $B$ denote signal and combinatorial background,
respectively,  and $a_{LL}$ is the analyzing power - the ratio of
spin-dependent and spin-independent cross sections in the PGF
process and is given as a function of photon as well as gluon
kinematics. The photon kinematics can be fully reconstructed based
on the incoming and scattered muons but the gluon part cannot be
reconstructed because only one charmed meson is measured. From MC
studies it was shown that the knowledge of the kinematics of only
one charmed meson can be used to reconstruct  the analyzing power
approximately. The parametrization of $a_{LL}$ was found by a
neural network trained using MC sample generated by the AROMA
generator and reconstructed as for real data.
Figure~\ref{fig:charm} shows the value of $\frac{\Delta G}{G}$ as
a function of year of data taking separately for $D^*$-tagged and
un-tagged events. Combining data from 2002-2004 we obtained the
following preliminary COMPASS result for the gluon polarization
from the open charm channels: $\frac{\Delta G}{G} = -0.57 \pm 0.41
(stat.)$ at $x_G \simeq 0.15$ and with the scale $\mu^2 \simeq 13
$ (GeV/c)$^2$. The systematic error studies are ongoing. We expect
the systematical uncertainty of the result to be smaller than
statistical error.

\section{$\frac{\Delta G}{G}$ from two high-$p_T$ hadrons}

\begin{figure}[t]
\begin{center}
\includegraphics[width=0.9\hsize]{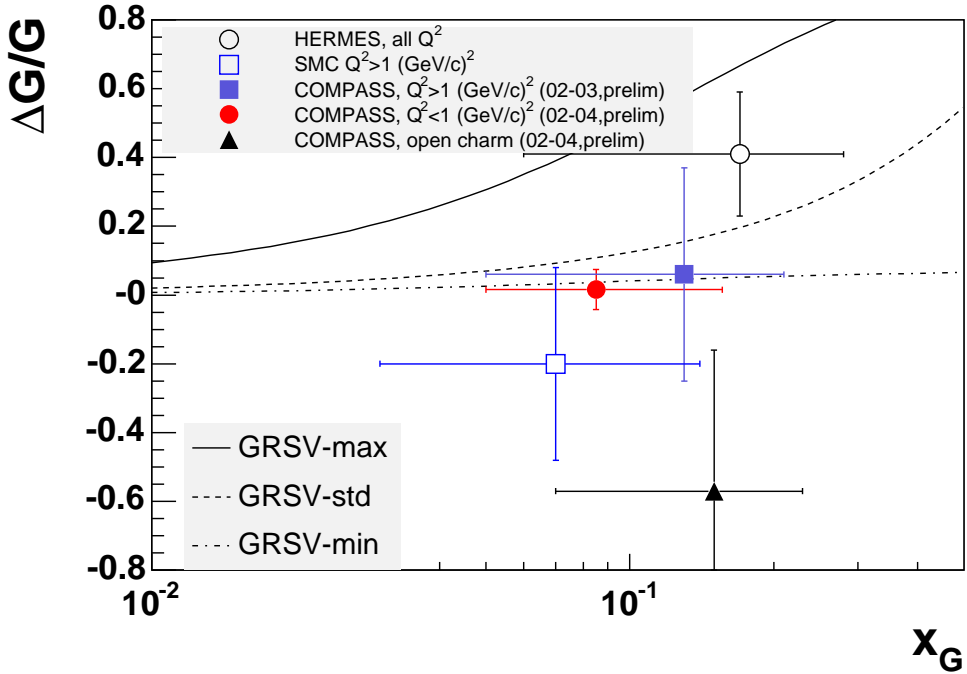}
\caption{Comparison of the $\frac{\Delta G}{G}$ measurement from
COMPASS, SMC and HERMES.}
\end{center}
\label{fig:all}
\end{figure}

Two parallel high-$p_T$ hadrons analyzes  are going on: the
so-called quasi-real photoproduction ($Q^2<1$(GeV/c)$^2$) of
high-$p_T$ hadron pairs  and the high-$p_T$ hadrons analysis with
high $Q^2$ ($Q^2>1$(GeV/c)$^2$). The reason for performing the
analysis in the two kinematical regions separately is that in the
two cases different background processes are contributing.
Corrections for this background have to be taken from MC
simulations and therefore very good agreement between data and MC
is required. The helicity asymmetry for two high-$p_T$ hadrons is
expressed as follows:
\begin{equation}
A_{LL}= R_{PGF} a_{LL}\frac{\Delta G}{G} + A_{Bkg.}, \label{eq:pt}
\end{equation}
where again $a_{LL}$ is the analyzing power for PGF subprocess,
$R_{PGF}$ is a fraction of PGF processes (taken from MC) and
$A_{Bkg}$ denotes the asymmetry from different background
processes which contribute to the observed two hadron final state.
In the low $Q^2$ sample the complicated background, including
resolved photon contribution was simulated with the PYTHIA MC
generator. An important contribution to the systematic error is
related to unknown polarized distribution functions in the photon
(two scenarios: plus or minus maximal polarization were taken into
account).The detailed procedure, data/MC comparison and results
for $\frac{\Delta G}{G}$ obtained for 2002-2003 data have been
recently published~\cite{pt}. The preliminary result - including
the 2004 data is: $ \frac{\Delta G}{G} = 0.016\pm 0.058 (stat.)\pm
0.055 (syst.)$ at $x_G \simeq 0.085^{+0.07}_{-0.035}$ and scale
$\mu^2 \simeq 3$ (GeV/c)$^2$. The comparison of all COMPASS $
\frac{\Delta G}{G}$ results and the results from SMC \cite{smc}
and HERMES \cite{hermes} experiments is presented in
Figure~\ref{fig:all}.

A new result for the high $Q^2$ - including the 2004 data set is
expected soon. In contrast to low $Q^2$ the physical background is
dominated by the leading  and the QCD-Compton process. The LEPTO
MC generator is used in this analysis. Due to the fact that large
$Q^2$ guarantees a perturbative scale the $p_T$ cuts can be
released increasing statistics. The neural network approach
similar to the one used in the SMC analysis \cite{smc} is now
tested.  We expect the reduction of the
statistical error by the factor up to 4 including 2004 data.\\
At the end it is worth to note that NLO corrections are partially
taken into account by using parton showers in MC generators and
string type fragmentation functions. Although model (MC) dependent
- the obtained result for low $Q^2$ is the most precise estimation
of the $\frac{\Delta G}{G}$ from directly measured helicity
asymmetry.

\section{Conclusions}
The new measurements of the gluon polarization obtained from the
COMPASS experiment have been presented. The model-independent
direct measurement based on the open charm channel and the most
precise but model (MC) dependent result from two high-$p_T$ hadron
pairs analysis indicate that a small $\Delta G$ is preferred. The
Ellis-Jaffe sum rule seems to be violated if a large $\Delta G$ is
excluded. The small $\Delta G$ indicates the important role of
angular orbital momentum in nucleon spin decomposition in the
frame of parton model and perturbative QCD.

\end{document}